\documentclass[twocolumn,floatfix,altaffilletter,superscriptaddress,preprintnumbers,
tightenlines,showpacs,showkeys,nofootinbib]{revtex4-1}
\usepackage[utf8]{inputenc}
\usepackage[colorlinks=true,citecolor=blue,linkcolor=blue]{hyperref}
\usepackage[normalem]{ulem}
\usepackage{amsmath,amssymb, mathrsfs}
\usepackage{epsfig}
\usepackage{graphicx}               
\usepackage{url}
\usepackage{color}
\usepackage{slashed}
\usepackage{multirow}
\usepackage{placeins}
\usepackage[dvipsnames]{xcolor}
\usepackage{epstopdf}
\usepackage{soul}
\usepackage{tikz}
\usepackage[capitalise, english]{cleveref}
\usepackage{siunitx}
\usepackage{xspace}
\usepackage{booktabs}
\usetikzlibrary{trees}
\usetikzlibrary{decorations.pathmorphing}
\usetikzlibrary{decorations.markings}

\newcommand\myshade{80}
\colorlet{mylinkcolor}{ForestGreen}
\colorlet{mycitecolor}{Red}
\colorlet{myurlcolor}{violet}

\hypersetup{
	linkcolor  = mylinkcolor!\myshade!black,
	citecolor  = mycitecolor!\myshade!black,
	urlcolor   = myurlcolor!\myshade!black,
	colorlinks = true
}

\definecolor{jblue}{RGB}{20,50,100}
\definecolor{npurple}{RGB} {153, 51, 204}
\definecolor{wred}{RGB}{217,0,56}
\definecolor{white}{RGB}{255,255,255}

\definecolor{korange}{RGB}{235, 80,  43}
\definecolor{korange2}{RGB}{245, 100,  63}
\definecolor{kyelloworange}{RGB}{255, 210,  110}
\definecolor{kyelloworange2}{RGB}{240, 170,  90}
\definecolor{kred}{RGB}{204,  102, 153}
\definecolor{kpurple}{RGB}{153,  61, 190}
\definecolor{kpurplelight}{RGB}{213,  161, 230}

\definecolor{tobycolour}{rgb}{.5,.0,.5}

\DeclareSIUnit\year{yr}
\DeclareSIUnit\pc{pc}
\DeclareSIUnit\ergs{ergs}
\DeclareSIUnit\msun{\ensuremath{M_\odot}}
\allowdisplaybreaks

\makeatletter
\providecommand*{\diff}%
{\@ifnextchar^{\DIfF}{\DIfF^{}}}
\def\DIfF^#1{%
	\mathop{\mathrm{\mathstrut d}}%
	\nolimits^{#1}\gobblespace}
\def\gobblespace{%
	\futurelet\diffarg\opspace}
\def\opspace{%
	\let\DiffSpace\!%
	\ifx\diffarg(%
	\let\DiffSpace\relax
	\else
	\ifx\diffarg[%
	\let\DiffSpace\relax
	\else
	\ifx\diffarg\{%
	\let\DiffSpace\relax
	\fi\fi\fi\DiffSpace}


\begin{document}
	
	\title{Neutrino and Positron Constraints on Spinning Primordial Black Hole Dark Matter}
	
	\author{Basudeb Dasgupta}
	\email{bdasgupta@theory.tifr.res.in}
	\thanks{\scriptsize \!\!  \href{http://orcid.org/0000-0001-6714-0014}{orcid.org/0000-0001-6714-0014}}
	\affiliation{Tata Institute of Fundamental Research, Homi Bhabha
		Road, Mumbai 400005, India}
	
	\author{Ranjan Laha} 
	\email{ranjan.laha@cern.ch}
	\thanks{\scriptsize \!\!  \href{http://orcid.org/0000-0001-7104-5730}{orcid.org/0000-0001-7104-5730}}
	\affiliation{Theoretical Physics Department, CERN, 1211 Geneva, Switzerland}
	
	\author{Anupam Ray}
	\email{anupam.ray@theory.tifr.res.in}
	\thanks{\scriptsize \!\!  \href{http://orcid.org/0000-0001-8223-8239}{orcid.org/0000-0001-8223-8239}}
	\affiliation{Tata Institute of Fundamental Research, Homi Bhabha
		Road, Mumbai 400005, India}
	
	\date{\today}
	
	
	\begin{abstract}
		Primordial black holes can have substantial spin -- a fundamental property that has a strong effect on its evaporation rate.  We conduct a comprehensive study of the detectability of primordial black holes with non-negligible spin, via the searches for the neutrinos and positrons in the MeV energy range. Diffuse supernova neutrino background searches and observation of the 511 keV gamma-ray line from positrons in the Galactic center set competitive constraints. Spinning primordial black holes are probed up to a slightly higher mass range compared to non-spinning ones. Our constraint using neutrinos is slightly weaker than that due to the diffuse gamma-ray background, but complementary and robust. Our positron constraints are typically weaker in the lower mass range and stronger in the higher mass range for the spinning primordial black holes compared to the non-spinning ones. They are generally stronger than those derived from the diffuse gamma-ray measurements for primordial black holes having masses greater than a few $\times \, 10^{16}$g.
	\end{abstract}

	\maketitle
	\preprint{CERN-TH-2019-212, TIFR/TH/19-40}


	\emph{Introduction.--} Astrophysical observations provide unambiguous evidence of a non-relativistic, collision-less, and weakly interacting matter, known as dark matter (DM), constituting $\sim$ 26$\%$ of the total energy density of the Universe~\cite{Aghanim:2018eyx}.  Many well-motivated DM candidates have been proposed and decades of experimental searches conducted, yet the microscopic identity of DM remains unknown. One of the earliest proposed DM candidates is a population of primordial black holes (PBHs)~\cite{1966AZh....43..758Z,Hawking:1971ei,Carr:1974nx, Chapline:1975ojl, Meszaros:1975ef, Carr:1975qj}.
	
	There exist numerous observational constraints on the fraction of DM comprised of PBHs (\cite{Carr:2009jm,Boudaud:2018hqb,Niikura:2017zjd,Smyth:2019whb, Griest:2013esa, Allsman:2000kg, Tisserand:2006zx,Wyrzykowski:2011tr, Brandt:2016aco, Ali-Haimoud:2016mbv,Raidal:2017mfl,Authors:2019qbw,Zumalacarregui:2017qqd}, see recent reviews \,\cite{Carr:2020gox,Carr:2020xqk,Green:2020jor} and references therein), however, there still exists parameter space where PBHs can form all of the DM\,\cite{Katz:2018zrn, Montero-Camacho:2019jte, Smyth:2019whb}.  Multiple ideas have also been proposed in order to probe PBHs in various mass ranges\,\cite{Jung:2019fcs, Bai:2018bej,Munoz:2016tmg,Laha:2018zav}. Given this increased scrutiny of PBHs (which started after the direct detection of gravitational waves\,\cite{Abbott:2016blz, LIGOScientific:2018jsj} and the subsequent proposal that these BHs are primordial in nature\,\cite{Bird:2016dcv, Cholis:2016kqi, Clesse:2016vqa, Sasaki:2016jop}), it is obvious to ask if we have explored all the properties of BHs in our searches for PBHs.  Typically, it has been assumed that PBHs have low spins\,\cite{Chiba:2017rvs, Mirbabayi:2019uph, DeLuca:2019buf}, however, there exist viable cosmological scenarios where PBHs are born with a high spin\,\cite{Khlopov:1980mg, Harada:2017fjm, Cotner:2018vug, Kokubu:2018fxy, He:2019cdb, Bai:2019zcd, Cotner:2019ykd, Arbey:2019jmj}.  Angular momentum is a fundamental property of BHs and it is crucial to explore its implications\,\cite{Dong:2015yjs, Arbey:2019vqx, Arbey:2019mbc, Kuhnel:2019zbc}.  Here, we study the impact of angular momentum on the observability of PBHs.  
	
	PBHs with masses $\lesssim 10^{-16}\,{\rm M}_\odot$ can be discovered via the observation of particles produced through Hawking radiation.  The lifetime of PBHs with masses less than $\ 2.5 \times 10^{-19}\, {\rm M}_\odot$ ($\ 3.5 \times 10^{-19} \, {\rm M}_\odot$) for non-rotating (maximally rotating) black holes is less than the age of the Universe and it cannot contribute to the DM density\,\cite{Page:1976df, Page:1976ki, Page:1977um, Taylor:1998dk}.  The leading constraints on low-mass PBHs arise from the observation of photons\,\cite{Carr:2009jm, Ballesteros:2019exr,Laha:2020ivk}, cosmic rays\,\cite{Boudaud:2018hqb}, and the 511 keV gamma-ray line\,\cite{DeRocco:2019fjq,Laha:2019ssq}.  Astrophysical observations of neutrinos have been used to constrain particle DM\,\cite{Dasgupta:2012bd, Murase:2015gea} and here we study its implications for PBHs.  Earlier analyses of the positron and the neutrino observations have focused on non-spinning PBHs\,\cite{Okele:1980kwj, okeke1980primary, 1991ApJ...371..447M, Bambi:2008kx, Bugaev:2000bz, Bugaev:2002yk, Bugaev:2002yt}.  Using the latest experimental inputs, we thoroughly investigate the constraints on spinning and non-spinning PBHs.
	
	The origin of the 511 keV gamma-ray line from the Galactic Center (GC) is one of the enduring mysteries of astrophysics~\cite{1973ApJ...184..103J, Knodlseder:2005yq, Prantzos:2010wi, Siegert:2015knp}.  Many models have been proposed to explain this observation\,\cite{Alexis:2014rba, Totani:2006zx,Sizun:2006uh,2017NatAs...1E.135C, Venter:2015gga, Prantzos:2005pz, Weidenspointner:2008zz, Bisnovatyi-Kogan:2016dgr, Farzan:2017hol}, yet none are confirmed.  Our constraints are agnostic of these models and robust. The diffuse supernova neutrino background (DSNB) is the accumulation of all neutrinos emitted by core-collapse supernovae over the history of the Universe\,\cite{Ando:2004hc, Beacom:2010kk, Lunardini:2010ab}.  The current upper limits on the $\bar{\nu}_e$ flavor of DSNB are due to the observations in {Super-Kamiokande}\,\cite{Bays:2011si}, {KamLAND}\,\cite{Collaboration:2011jza}, and Borexino\,\cite{Agostini:2019yuq}. 
	
	We probe the DM fraction of non-spinning and spinning PBHs by considering neutrino and positron emission and possible detection by {Super-Kamiokande} (Super-K) and {INTEGRAL}, respectively. By considering a log-normal mass distribution, we can probe further into a previously unexplored mass window. At low masses, our results constrain the possibility that PBHs make up all the DM.

	\emph{Methods \& Results.--} PBHs can have wide range of masses depending on its formation time~\cite{Carr:2009jm}. An uncharged rotating BH radiates with a temperature~\cite{Hawking:1974sw, Page:1976df, Page:1976ki, MacGibbon:1990zk, MacGibbon:2007yq}
	\begin{equation}
	T_{\rm{PBH}} = \frac{1}{4 \pi G_{N} M_{\rm{PBH}}}  \frac{\sqrt{1-a_*^2}}{1+\sqrt{1-a_*^2}}\,,
	\label{eq:PBH temperature}
	\end{equation}
	where $M_{\rm PBH}$ denotes the mass of the PBH, $G_N$ is the gravitational constant, $a_* \equiv J/(G_N M_{\rm PBH}^2)$ is the reduced spin parameter, and $J$ is the angular momentum of the PBH.  For a given PBH mass, the temperature can vary by orders of magnitude as the PBH spin approaches its extremal value, $a_* \rightarrow$ 1.

	\begin{figure*}[!t]
		\centering
		\includegraphics[angle=0.0,width=0.32\textwidth]{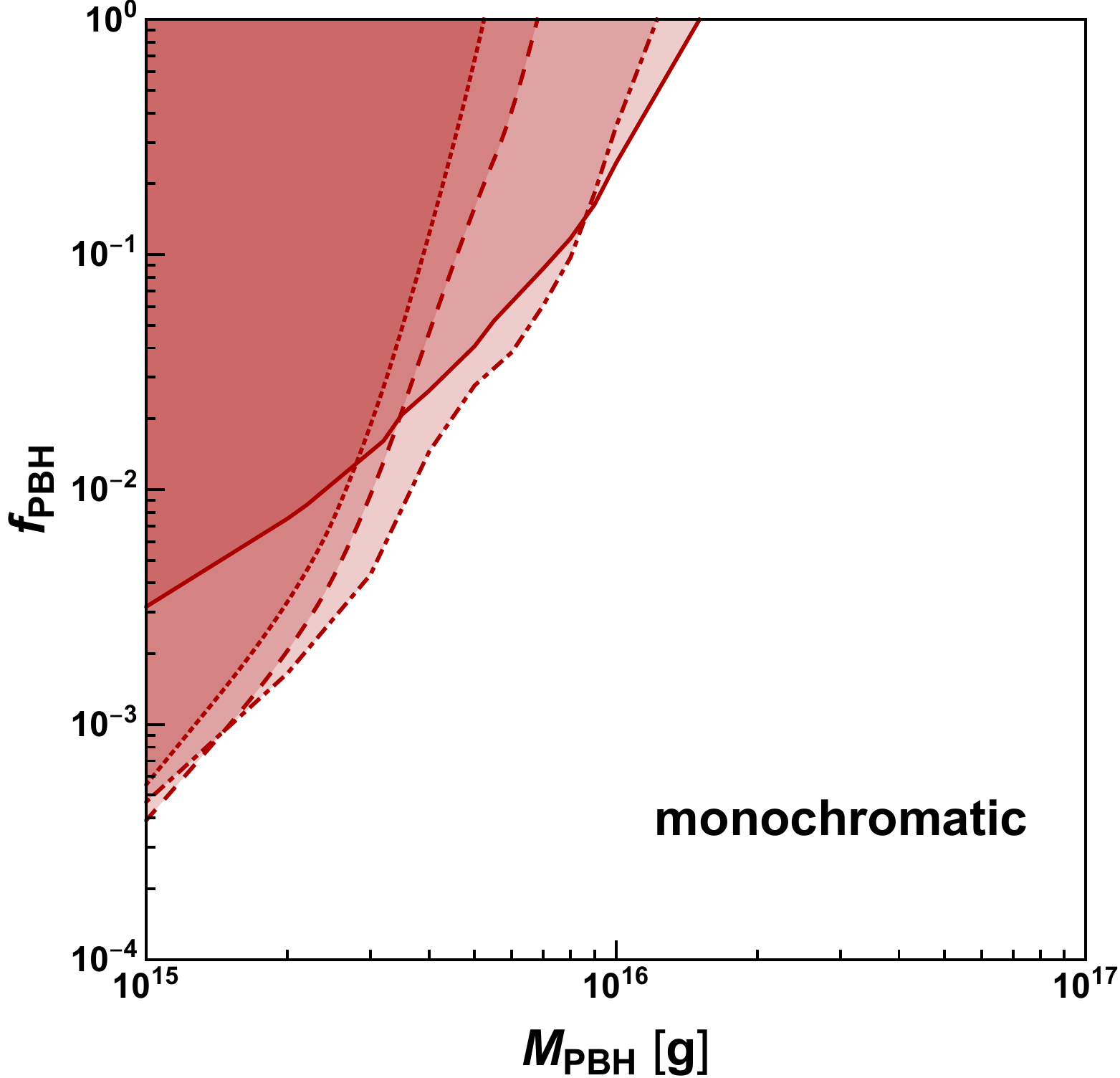}
		\includegraphics[angle=0.0,width=0.32\textwidth]{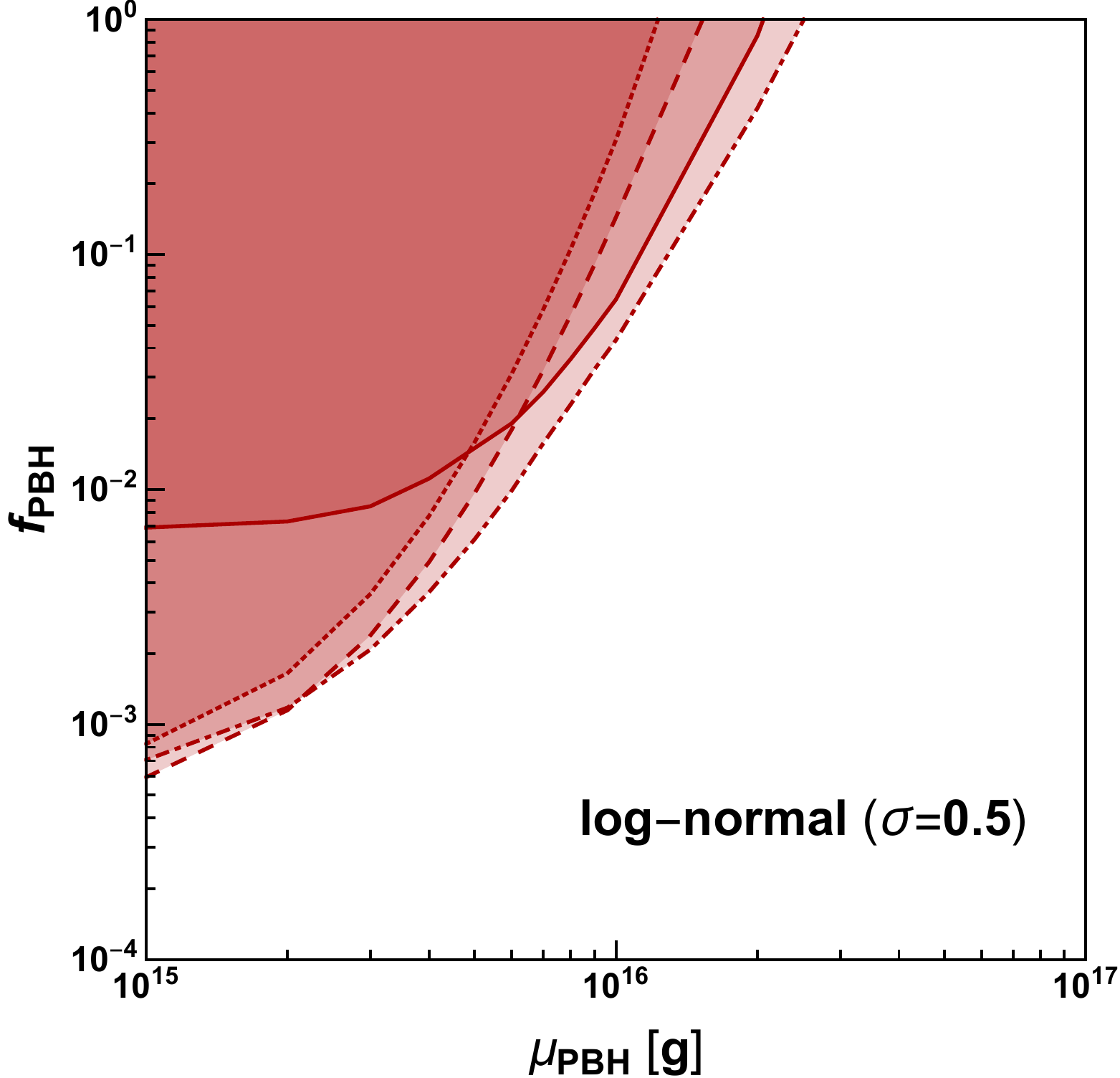}
		\includegraphics[angle=0.0,width=0.32\textwidth]{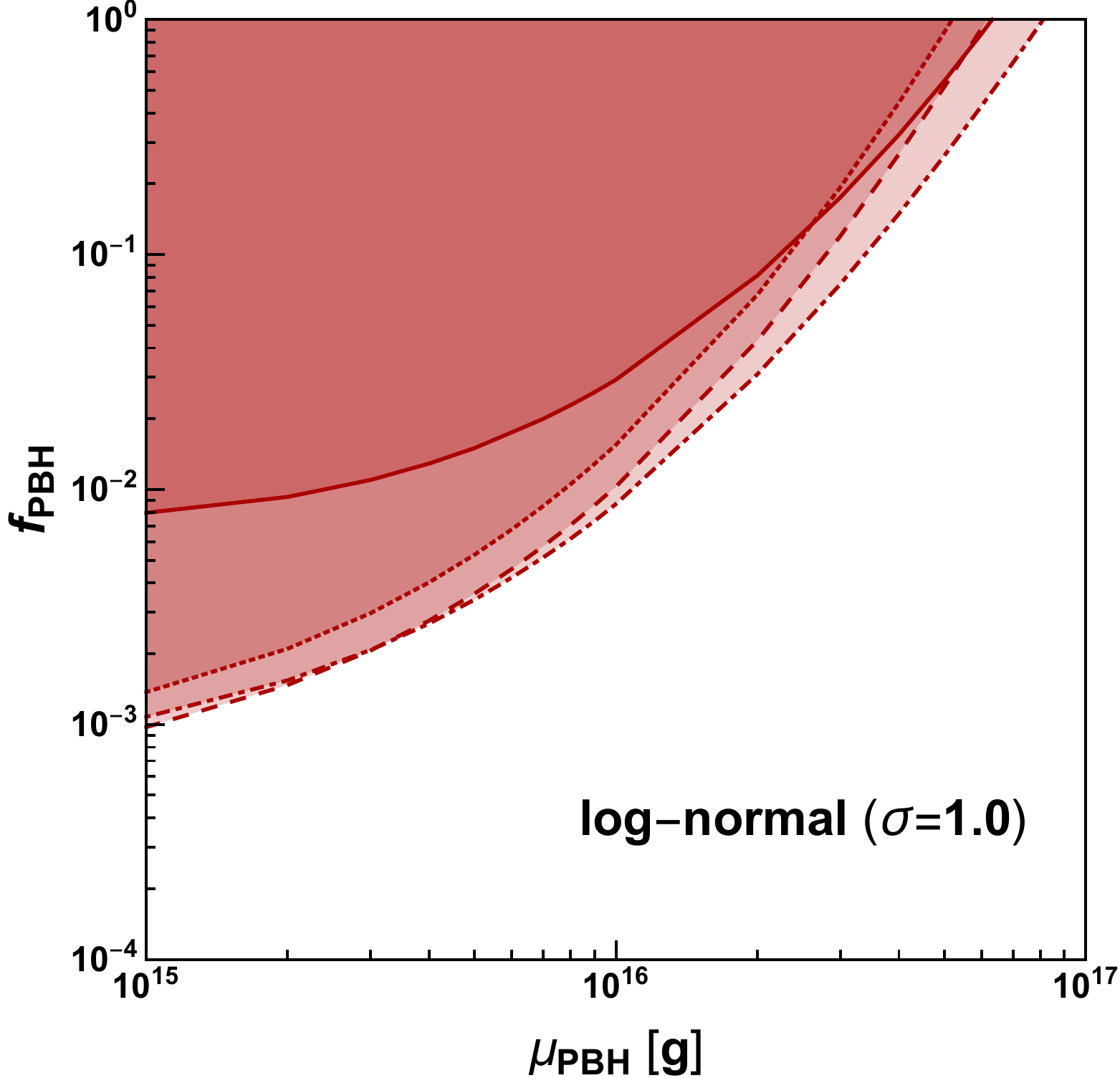}
		\caption{Upper limit on dark matter fraction of PBHs, $f_{\rm PBH}$, from DSNB searches at Super-Kamiokande. The left, middle, and right panel corresponds to a monochromatic PBH mass function and log-normal PBH mass functions with $\sigma$ = 0.5 and 1.0, respectively. In each plot, four different lines correspond to four different reduced spin parameters $(a_*=0,0.5,0.9,0.9999)$ of PBHs. Tiny dotted, dashed, dot-dashed, and solid lines correspond to $a_*=0$, $a_*=0.5$, $a_*=0.9$, $a_*=0.9999$, respectively.  These constraints are derived using an NFW dark matter profile of the Milky Way.}
		\label{fig: neutrino limits}
	\end{figure*}
	
	The number of emitted particles with spin $s$ in the energy interval $E$ and $E+dE$ and in a time interval $dt$ from a PBH is
	\begin{equation}
	\frac{d^2N}{dEdt} = \frac{1}{2\pi} \frac{\Gamma_s(E,M_{\rm{PBH}},a_*)}{{\exp}\left[{E'}/{T_{\rm{PBH}}} \right]-(-1)^{2s}} \,,
	\label{eq:Differential energy distribution}
	\end{equation} 
	where $\Gamma_s$ is the graybody factor~\cite{Page:1976df, Page:1976ki, Page:1977um, MacGibbon:1990zk, MacGibbon:1991tj, MacGibbon:2007yq, Arbey:2019vqx} and $E'$ is the total energy of the emitted species including the rotational velocity.  For the rest of our analysis, we will use {\tt BlackHawk} to compute the spectra of the emitted particle~\cite{Arbey:2019mbc}.  We have checked this emission rate using the semi-analytical formulas from Ref.\,\cite{Page:1976df, Page:1976ki, Page:1977um, MacGibbon:1990zk}. 
	
	As the temperature of a PBH becomes comparable to the energy of a particle, such a particle is emitted in significant numbers\,\cite{Page:1976df, Page:1976ki, Page:1977um}.  We first focus on the emission of neutrinos from PBHs.  In order to derive bounds from neutrinos, we need to take into account the Galactic and extragalactic contribution of PBHs.  The Galactic contribution is given by  
	\begin{equation}
	F_{\rm Gal} = \int \frac{d\Omega}{4\pi}  \int dE \frac{d^2N}{dE dt} \int dl\,\frac{f_{\rm PBH}\,\rho_{\rm MW}\left[r(l,\psi)\right]}{M_{\rm{PBH}}}\,,
	\label{eq: Galactic neutrino contribution}
	\end{equation}
	where $r$ is the galactocentric distance, $\rho_{\rm MW}(r)$ denotes the DM profile of the Milky Way (MW), $l$ is the distance from the observer, $\psi$ is the angle between the line of sight and the observer, $\Omega$ is the solid angle under consideration, and the fraction of DM composed of PBHs is denoted by $f_{\rm PBH}$ .  The upper limit of the line of sight integral, $l_{\rm max}$, depends on the MW halo size and $\psi$\,\cite{Ng:2013xha}.  We use the NFW and the isothermal DM profiles using the parametrization in Ref.\,\cite{Ng:2013xha}.  For the extragalactic contribution, the differential flux integrated over the full sky is~\cite{Carr:2009jm,Arbey:2019vqx}
	\begin{equation}
	F_{\rm EG} =  \int\int dt\,d\tilde{E}\,\left[1+z(t)\right] \frac{f_{\rm{PBH}} \rho_{\rm DM}}{M_{\rm{PBH}}} \frac{d^2N}{dE dt}\Bigr\rvert_{E = [1+z(t)]\tilde{E}}\,,
	\label{eq:extragalactic contribution}
	\end{equation}
	where the time integral runs from  $t_{\rm min}$= 1\,s, the neutrino decoupling time, to $t_{\rm max}$, the smaller of the PBH lifetime and age of the Universe. Although the ultralight PBHs  are formed much earlier than the neutrino decoupling time, we have taken it as a lower limit of the time integral $(t_{\rm min})$,  because neutrinos emitted from PBHs can free stream after neutrino decoupling. Note that, changing this lower limit to smaller values has  very little effect on the corresponding upper limit.  The average DM density of the Universe at  the present time is denoted by $\rho_{\rm DM}$.  We use the cosmological parameters determined by the Planck observations\,\cite{Aghanim:2018eyx}.  
	
	In addition to a monochromatic mass function for PBHs, we also consider a log-normal mass function, as predicted by many inflation models:
	\begin{equation}
	\dfrac{dN_{\rm PBH}}{dM_{\rm PBH}}= \frac{1}{\sqrt{2\pi}\sigma M_{\rm{PBH}}}\, \exp\left[- \dfrac{ {\rm ln}^2 \left(M_{\rm PBH}/\mu_{\rm PBH}\right)}{2 \sigma^2} \right]\,,
	\label{eq: log-normal mass distribution}
	\end{equation} 
	where $\mu_{\rm{PBH}}$ and $\sigma$ are the average mass and width of the distribution, respectively.
	
	The upper limit on $f_{\rm PBH}$ is obtained by comparing the total Galactic and extragalactic flux due to PBHs, with the current upper limit on the neutrino flux from different experiments. We find that neutrino experiments looking for the DSNB are able to set interesting constraints.  Neutrinos are emitted as mass eigenstates during PBH evaporation\,\cite{Lunardini:2019zob}. So, for $T_{\rm PBH}\gg m_{\nu}$, the $\bar{\nu}_{e}$ flux is approximately equal to that of any one of the mass eigenstates. Current upper limits on the DSNB flux are 2.9 $\bar{\nu}_e$\,$\rm{cm}^{-2}\,s^{-1}$ (139  $\bar{\nu}_e$ $\rm{cm}^{-2}\,s^{-1}$) in the energy $(E_{\bar{\nu}_e})$ range of  17.3 to 91.3 MeV (8.3 to 31.8 MeV) respectively~\cite{Bays:2011si, Collaboration:2011jza}.  We find that  the Super-Kamiokande  and the KamLAND data help us probe the physical region of $f_{\rm PBH} < 1$.  We only show the upper limit obtained using the Super-Kamiokande data, as it is stronger at all PBH masses we consider.

	\begin{figure*}
		\centering
		\includegraphics[angle=0.0,width=0.32\textwidth]{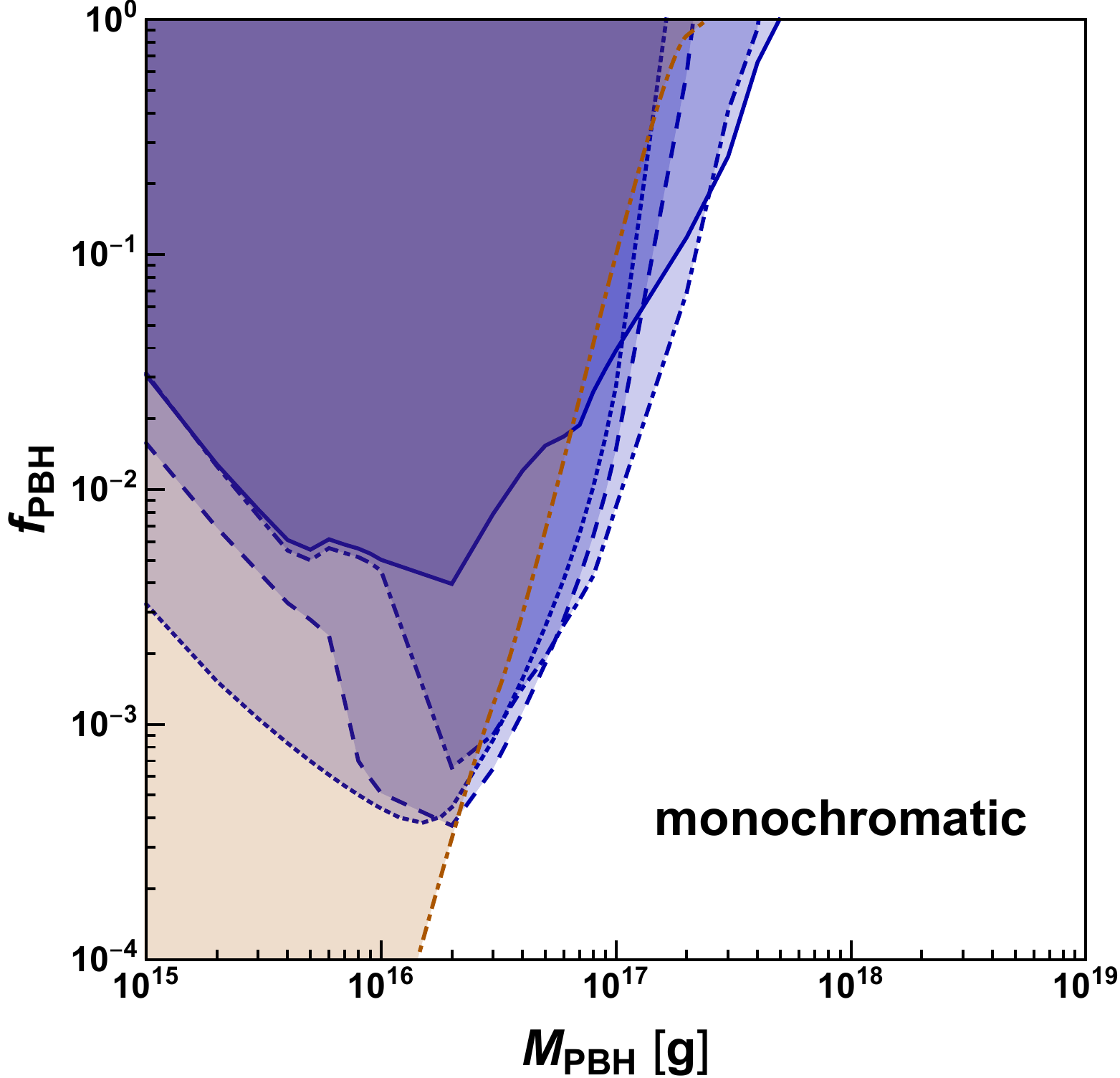}
		\includegraphics[angle=0.0,width=0.32\textwidth]{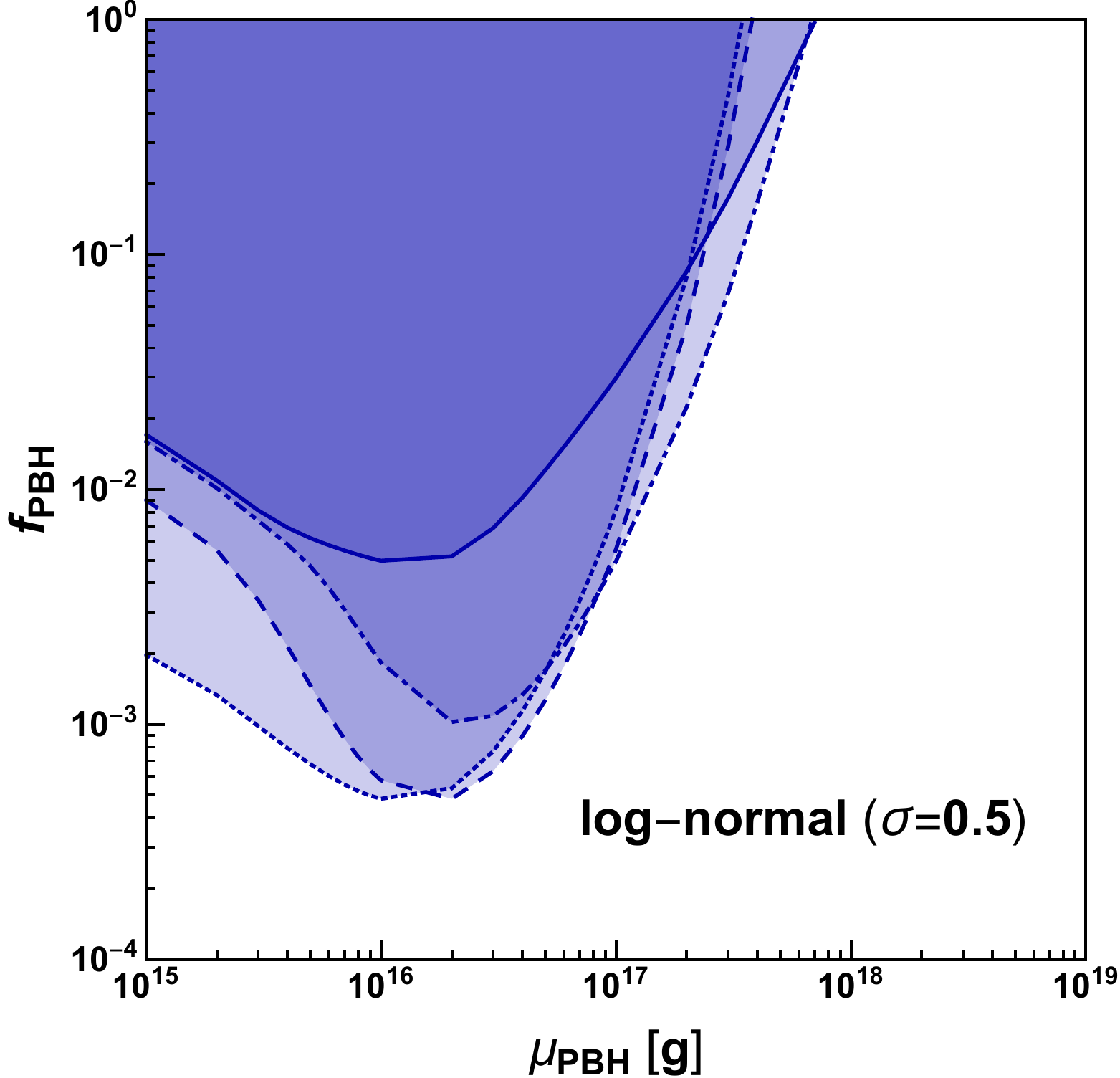}
		\includegraphics[angle=0.0,width=0.32\textwidth]{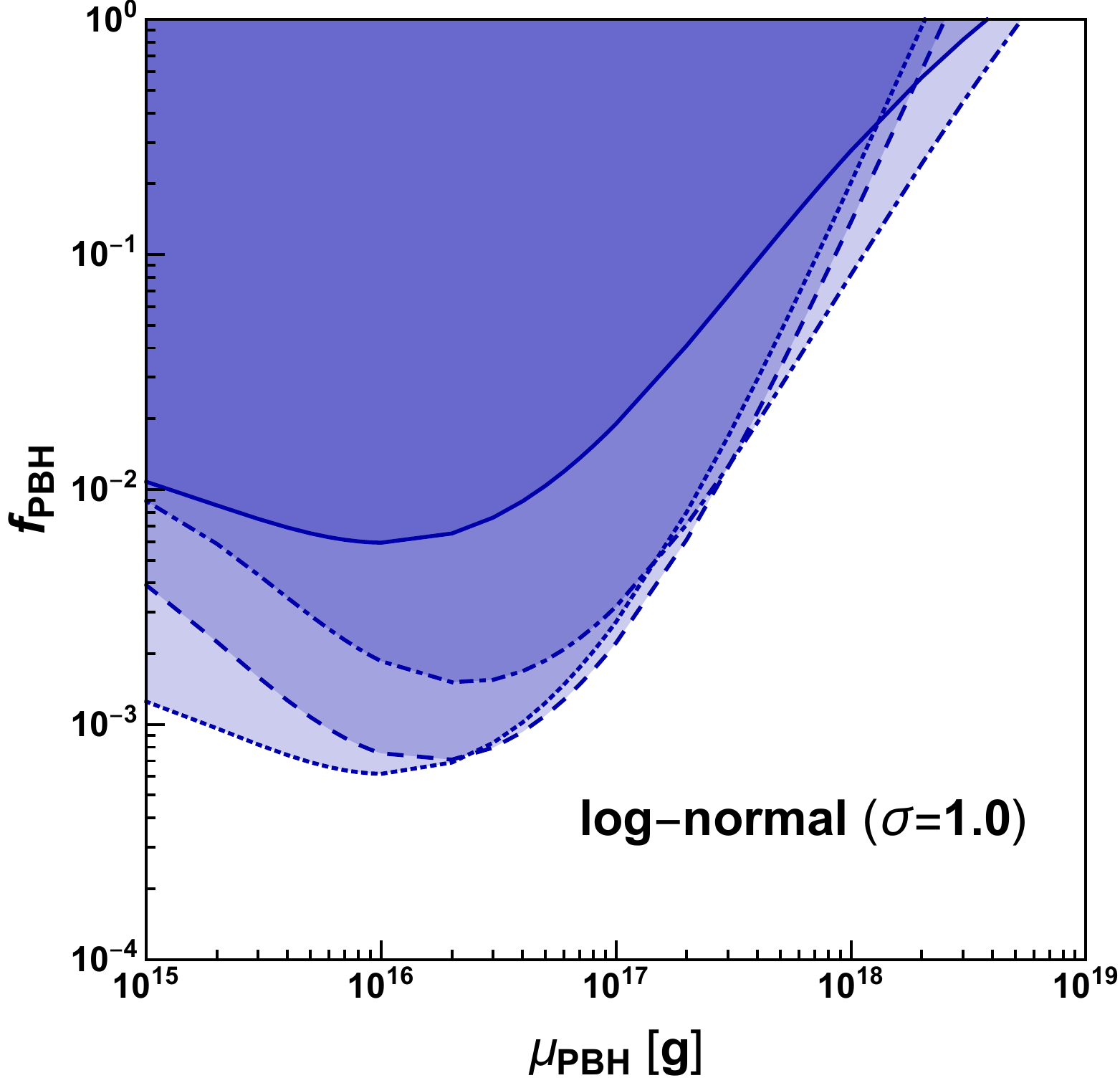}
		\caption{Upper limit on dark matter fraction of PBHs, $f_{\rm PBH}$, from INTEGRAL 511 keV gamma-ray line measurement. The left, middle, and right panels correspond to a monochromatic PBH mass function and log-normal PBH mass functions with $\sigma$ = 0.5 and 1.0, respectively. In each plot, four different lines correspond to four different reduced spin parameters $(a_*=0,0.5,0.9,0.9999)$ of PBHs. Tiny dotted, dashed, dot-dashed, and solid lines correspond to $a_*=0$, $a_*=0.5$, $a_*=0.9$, $a_*=0.9999$, respectively.  These constraints are derived using an NFW dark matter profile of the Milky Way and assume that 80\% of positrons within 3.5 kpc of the Galactic center annihilate to produce the 511 keV signal. These constraints are derived using an NFW dark matter profile of the Milky Way and assume that 80\% of positrons within 3.5 kpc of the Galactic Center annihilate to produce the 511 keV signal. In the left panel, the orange line shows the related limit derived from gamma-ray observations\,\cite{Arbey:2019vqx} for $a_*=0.9$. Note that or constraints are more stringent at high mass.}
		\label{fig: positron constraints}
	\end{figure*}

	Fig.\,\ref{fig: neutrino limits} shows the upper limits on $f_{\rm PBH}$ that can be derived from DSNB searches, for various PBH mass distributions and spins.  The left panel shows the constraints for the monochromatic mass distribution, whereas the middle and the right panels show the constraints for a log-normal distribution with $\sigma$ = 0.5 and 1, respectively.  For all these cases, we choose $a_*$ = 0, 0.5, 0.9, and 0.9999 and the NFW profile to derive our limits.  Since spinning BHs evaporate faster\,\cite{Page:1976df, Page:1976ki, Page:1977um}, the limits for $a_*$ = 0.5 and 0.9 are stronger than the non-spinning cases for the three mass distributions.  For $a_*$ = 0.9999, we find that the constraints are typically weaker than for $a_*$ = 0.9.  This can be understood by the rapid decrease in $T_{\rm PBH}$ as $a_* \rightarrow$ 1 for a fixed value of $M_{\rm PBH}$.  Because of the much smaller temperature, there are fewer neutrinos in the relevant energy interval, giving a weaker constraint.  This constraint is very robust and conservative. It depends on the total DM mass of the MW, but not on the DM profile of the MW. Unlike the positron derived constraint, neutrino derived constraint is minimally sensitive to propagation related uncertainties. These  can be due to matter effect and  uncertainties in the oscillation parameters.
	
	For the detection of much lower energy neutrinos, PTOLEMY is a proposed experiment with the capability to detect the cosmic neutrino background~\cite{Weinberg:1962zza, Cocco:2007za, Long:2014zva, Betti:2019ouf}.  We found that the event rate of low energy neutrinos coming from the PBH evaporation is incredibly small in PTOLEMY and thus it will not be able to set useful limits\,\cite{Lunardini:2019zob}.
	
	The constraints from the GC positrons are derived following Ref.\,\cite{Laha:2019ssq}.  Given the plethora of astrophysical models to explain the GC 511 keV line, we derive the most conservative bound by simply requiring that the number of positrons injected via PBH evaporation is smaller than the number of positrons required to explain the observed 511 keV line.  The major uncertainty in this technique arises from the unknown propagation distance of positrons in the GC\,\cite{Higdon:2007fu,Alexis:2014rba,Panther:2018xvc}.

	The observed flux of 511 keV photons implies that the total production rate of positrons within the Galactic bulge is $ \sim 10^{50} \ \rm{yr}^{-1}$~\cite{Fuller:2018ttb,Prantzos:2010wi,Siegert:2015knp}. The limit on the PBH fraction of DM $(f_{\rm{PBH}})$ is obtained by requiring that positron injection rate from PBHs obeys this constraint:
	\begin{equation}
	f_{\rm{PBH}} \leq \frac{10^{50}\ \rm{yr}^{-1}}{\int dE \int d M_{\rm{PBH}} \frac{dN_{{\rm PBH}}}{dM_{\rm PBH}} \frac{d^2N}{dEdt}\int \frac{d^3r\,\rho_{\rm MW}(r)}{M_{\rm{PBH}}}} \, . 
	\label{eq: INTEGRAL upper limit}
	\end{equation}
	The energy interval in the above expression runs from 0.511\,MeV to 3\,MeV~\cite{Beacom:2005qv}.  A careful astrophysical modeling of the sources can improve this limit by an order of magnitude~\cite{Laha:2019ssq}.  In order to account for the propagation uncertainty of positrons, we consider two different cases: $(i)$ all positrons injected within 1.5 kpc of the GC annihilate to produce the 511 keV signal and $(ii)$ 80\% of positrons injected within 3.5 kpc of the GC annihilate to produce the 511 keV signal~\cite{Higdon:2007fu,Alexis:2014rba,Panther:2018xvc}.
	
	Fig.\,\ref{fig: positron constraints} shows the upper limit on $f_{\rm PBH}$, from the GC positron observation for various PBH mass functions and reduced spin parameters.  The left, middle, and right panels display the constraints for the monochromatic PBH mass function and the log-normal mass distribution with $\sigma$ = 0.5 and 1, respectively.  In each panel, constraints are shown for four different reduced spin parameters, $a_*$ = 0, 0.5, 0.9, and 0.9999.  Because of the semi-relativistic nature of the positrons, the constraints for $a_*$ = 0 is the most stringent, especially at low PBH masses.  One can analytically show PBH spin can lead to $T_{\rm{PBH}}<E_{e}$, suppressing positron emission.  Because of this suppressed emission, the constraints on $f_{\rm PBH}$ weaken in parts of the parameter space. Note that, the constraints derived from the INTEGRAL observations are stronger than those derived from the diffuse gamma-ray measurements, especially at PBH masses $\gtrsim$ few $\times$ 10$^{16}$ g. See Fig.\,\ref{fig: Comparison gamma INTEGRAL} in the Supplemental Material (SM) for a more detailed comparison. 
	
	\begin{figure}
		\includegraphics[angle=0.0,width=0.48\textwidth]{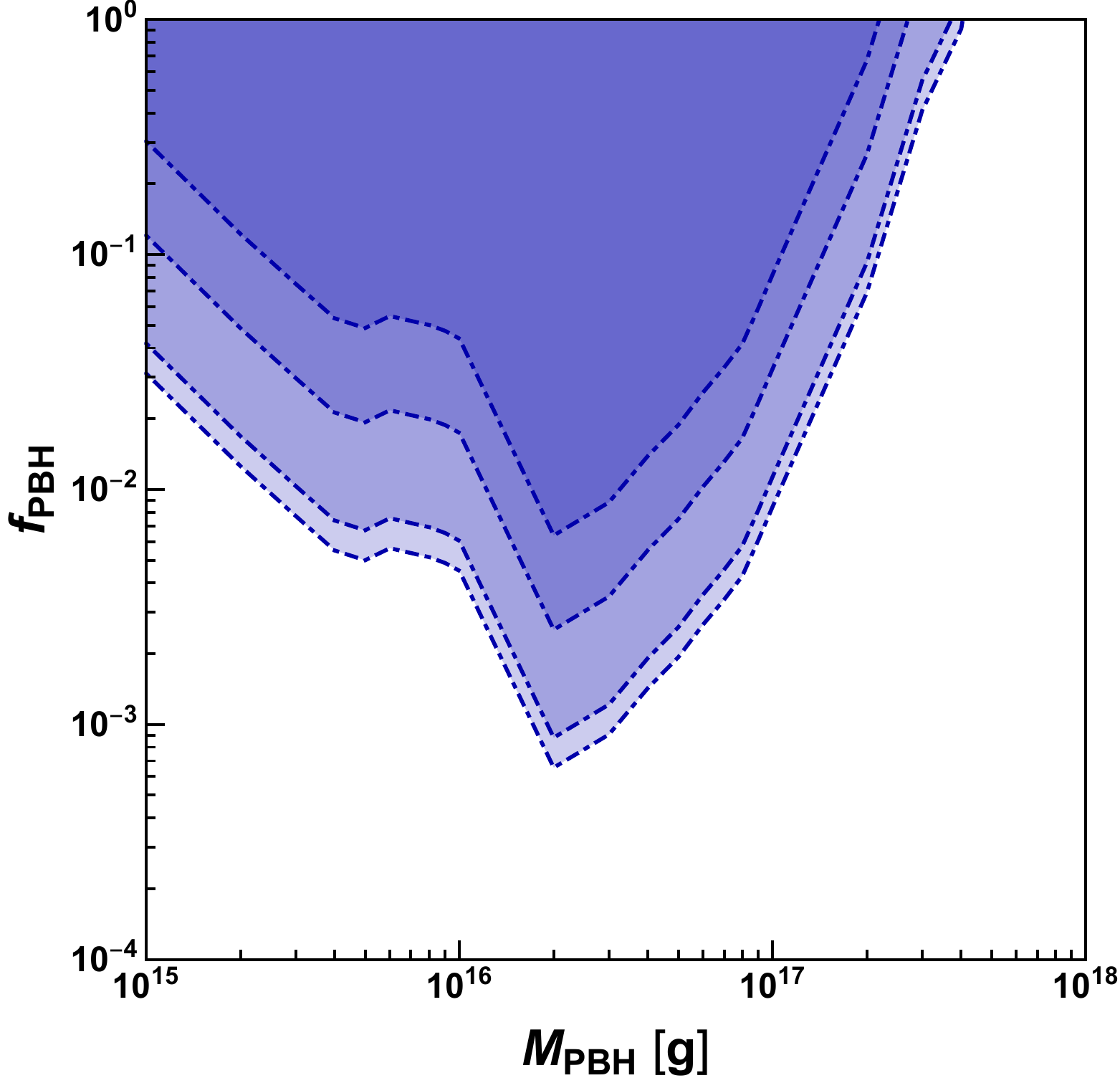}
		\caption{Variation in the upper limit on dark matter fraction of PBHs for monochromatic mass distribution from INTEGRAL 511 keV gamma-ray line measurement, due to dark matter density profiles and positron propagation. This plot considers a PBH with $a_*=0.9$.  The lines from top to bottom correspond to isothermal with 1.5 kpc, NFW with 1.5 kpc, isothermal with 3.5 kpc, and NFW with 3.5 kpc region of interest, respectively.}
		\label{fig: INTEGRAL variation}
	\end{figure}
	
	Fig.\,\ref{fig: INTEGRAL variation} shows the variation of the positron constraints for $a_*$ = 0.9, for different DM profiles and  propagation distance of low-energy positrons in the GC.  Since this variation is a multiplicative constant, as evident from Eq.\,(\ref{eq: INTEGRAL upper limit}), this uncertainty is the same for PBHs with different spins.  The strongest constraint arises when we consider that the DM profile is NFW and that 80\% of positrons injected within 3.5 kpc of the GC annihilate to produce the 511 keV signal.  The weakest constraint arises with the isothermal DM profile and a 1.5 kpc region of interest around the GC.

	\emph{Summary \& Outlook.--} Although PBHs had been written off as the dominant form of DM  several years ago, recent studies indicate that such a conclusion was premature. Our constraints represent a valuable contribution to the now ongoing, more careful, reappraisal of the situation. See Fig.\,\ref{fig: Global plot} in the SM for a bird's-eye view of available constraints. Angular momentum, a fundamental property of BHs, can drastically change the evaporation rate of a BH.  There has been a recent surge of interest in spinning PBHs and it is necessary to fully explore the parameter space of these exotic objects. Using DSNB searches and the INTEGRAL observations of the Galactic center 511 keV gamma-ray line, we probe the allowed parameter space of uncharged spinning PBHs.  We show that nonzero angular momentum of a PBH allows us to probe higher mass PBHs.  Our constraints using the DSNB (INTEGRAL) observations are weaker (stronger) than the existing limits from the diffuse gamma-ray background. Improved modeling of the GC positrons will also allow us to probe PBHs more comprehensively. These constraints depend on the underlying mass distribution~\cite{Carr:2016drx,Carr:2017jsz,Green:2016xgy,Kuhnel:2017pwq,Bellomo:2017zsr}, however, it is generally true that there exist multiple mass distributions for which PBHs can make up the entire DM density. Although our neutrino constraints are somewhat weaker, they are very robust to uncertainties in the DM density profile (depending only on the total DM mass) and to a variety of astrophysical uncertainties that are inevitably associated with photons or charged particles. Near-future loading of gadolinium in Super-Kamiokande and Hyper-Kamiokande will further enhance their capability to search for the DSNB\,\cite{Beacom:2003nk}. Possible DSNB detection, followed by a modeling of the stellar background, will greatly increase the prospect of PBH discovery via neutrinos. 
	
	\emph{Acknowledgments.--} We thank Jeremy Auffinger for help with the {\tt BlackHawk} package.  We are grateful to Jane MacGibbon for pointing us to relevant older literature. The work of B.D. is supported by the Department of Atomic Energy (Govt.\,\,of India) research project 12-R\&D-TFR-5.02-0200, the Department of Science and Technology (Govt.\,\,of India) through a Ramanujan Fellowship, and by the Max-Planck-Gesellschaft through a Max Planck Partner Group. R.L. thanks the CERN Theory Group for support.
	
\bibliographystyle{apsrev4-1}
\bibliography{ref.bib}

\clearpage
\newpage
\maketitle
\onecolumngrid
\begin{center}
\textbf{\large Supplemental Material} \\

\vspace{0.05in}

\textbf{\large Neutrino and Positron Constraints on Spinning Primordial Black Hole Dark Matter} \\

\vspace{0.07in}

{Basudeb Dasgupta, Ranjan Laha, and Anupam Ray}

\end{center}

	In this Supplementary Material we present comparisons of the positron constraints that we derived in the main text to other related constraints on primordial black hole dark matter.

\section{Comparison to other limits} 

\begin{figure}[!h]
	\includegraphics[angle=0.0,width=0.48\textwidth]{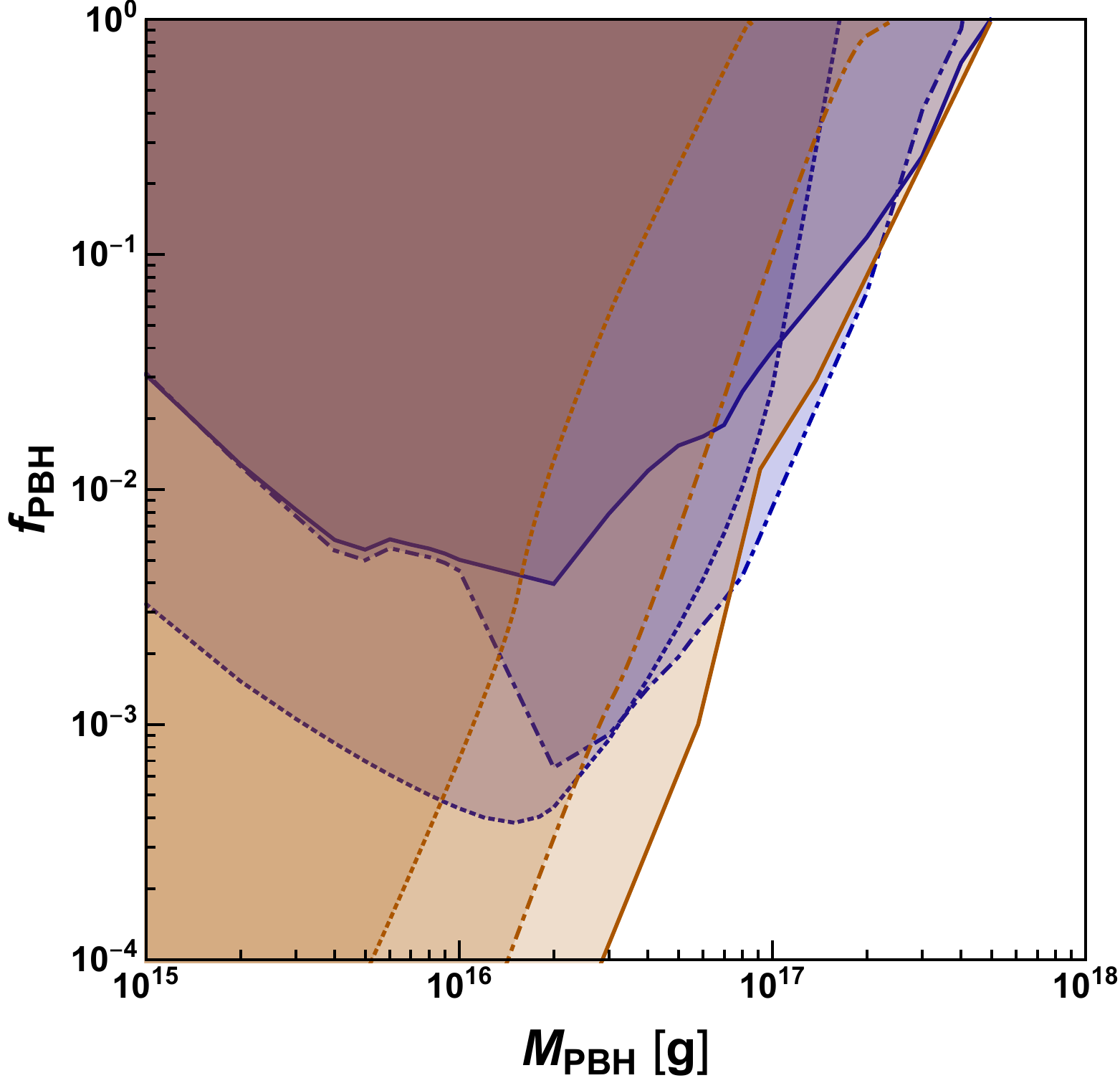}	\quad
	\includegraphics[angle=0.0,width=0.48\textwidth]{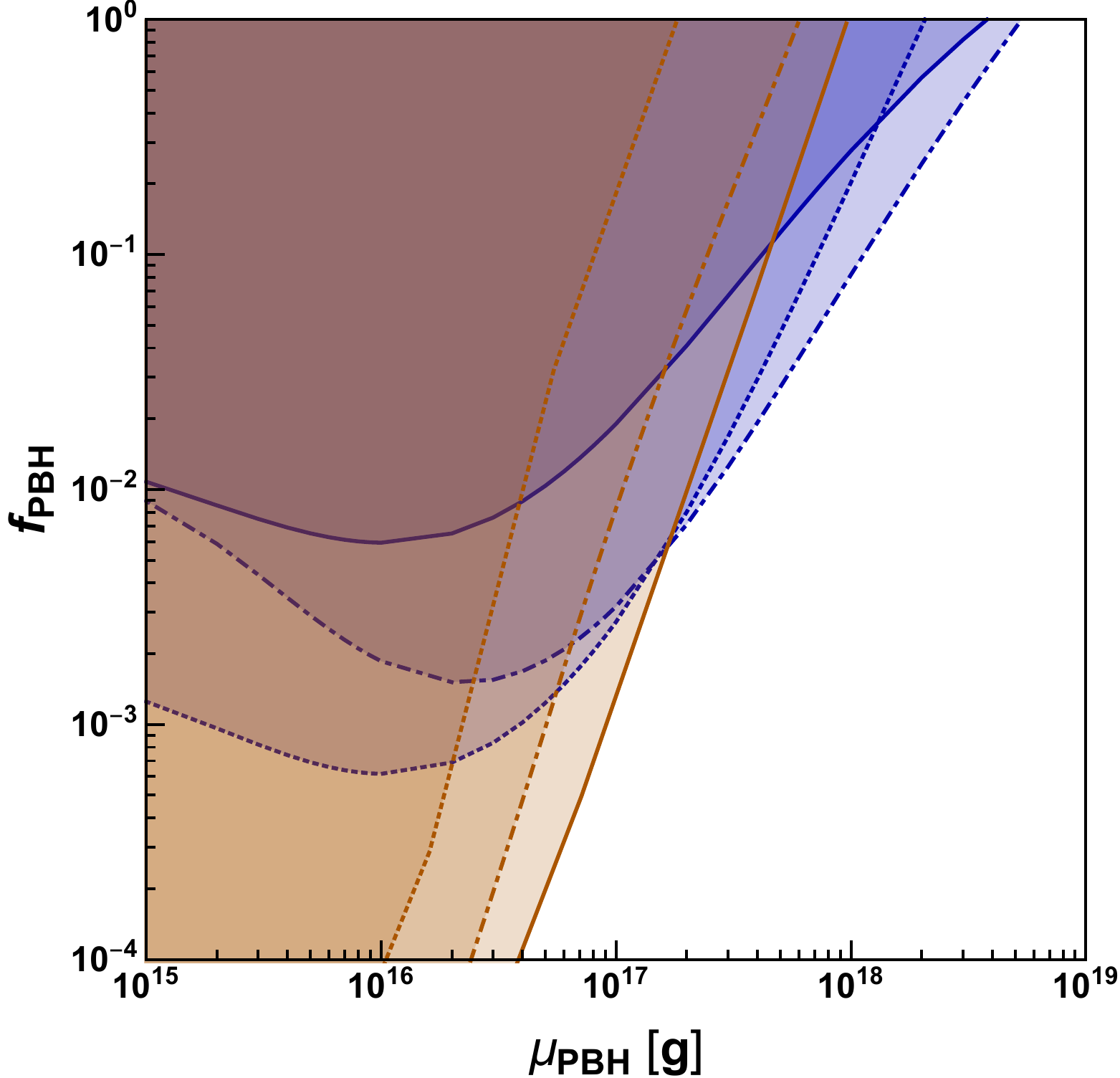}
	\caption{Comparison of the limits derived from gamma-ray observations\,\cite{Arbey:2019vqx} (in various shades of orange) with that derived in this work (in various shades of blue) using the INTEGRAL 511 keV gamma-ray line measurements.  For the latter, we have used an NFW dark matter profile and assumed that 80\% of positrons injected via PBH evaporation with 3.5 kpc of the Galactic center annihilate.  The line styles have the same meaning as in Fig.\,2 in the main text.  The left (right) panel corresponds to a monochromatic (log-normal distribution with $\sigma$ = 1) PBH mass distribution.}
	\label{fig: Comparison gamma INTEGRAL}
\end{figure}

Fig.\,\ref{fig: Comparison gamma INTEGRAL} shows the comparison of our limits from INTEGRAL observations with that derived from gamma-ray measurements, for the monochromatic PBH mass distribution (left panel) and the log-normal mass distribution with $\sigma$ = 1 (right panel).  The constraints derived from the INTEGRAL observations are stronger than those derived from the diffuse gamma-ray measurements, especially at PBH masses $\gtrsim$ few $\times$ 10$^{16}$ g. For example, for monochromatic mass distribution, the positron derived constraint on a PBH with mass $\sim 2 \times 10^{17}$ g and $a_*$ = 0.9  is approximately an order of magnitude stronger than the corresponding constraint from gamma-rays.  Similarly for the log-normal mass distribution, $\sigma$ = 1, the constraint on $\mu_{\rm PBH}$ $\sim$ 6 $\times$ 10$^{17}$ g and $a_*$ = 0.9 from positrons is about a factor of $\sim$ 25 stronger than the corresponding gamma-ray constraint. 

\begin{figure*}
	\centering
	\includegraphics[angle=0.0,width=0.98\textwidth]{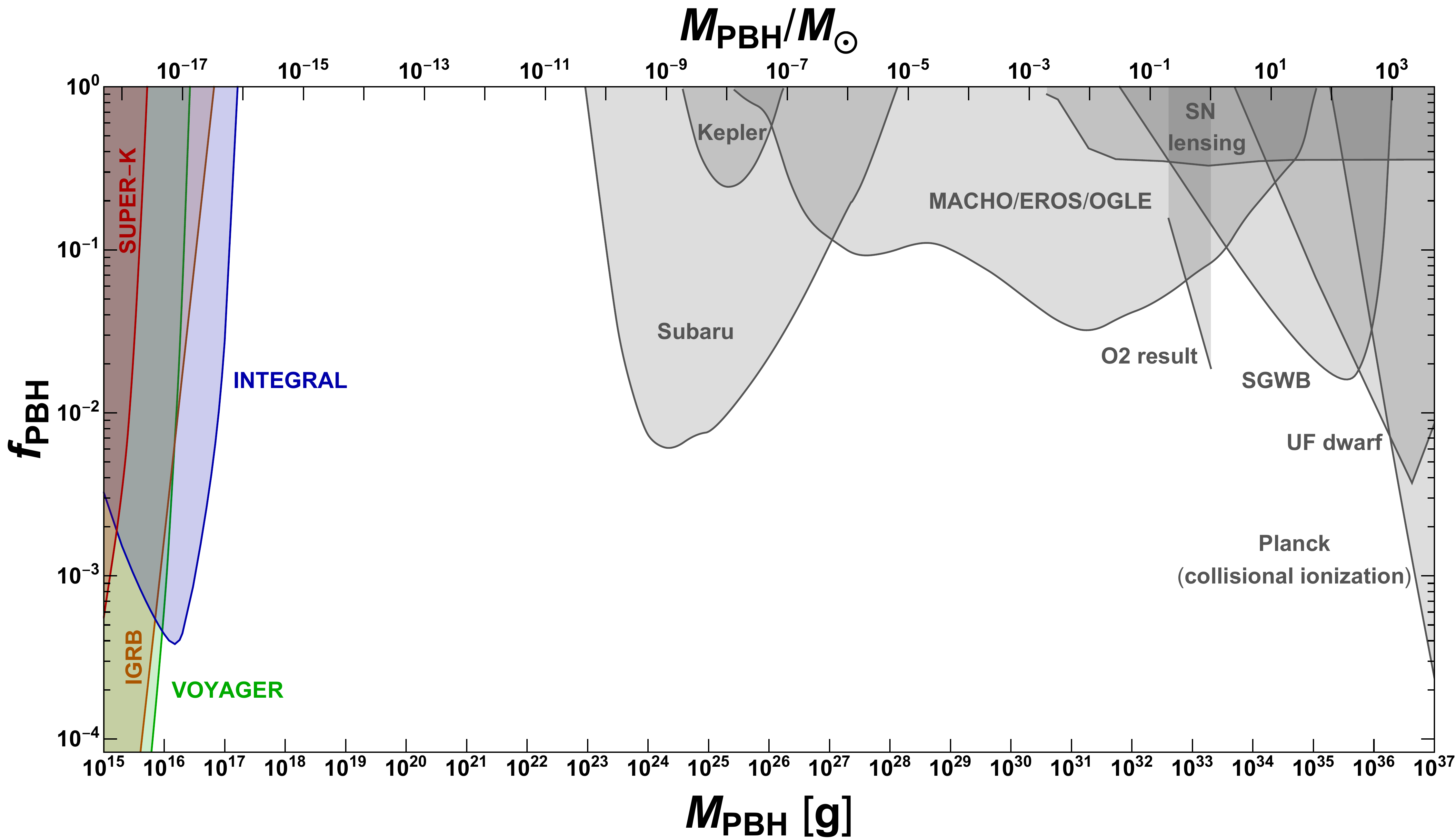}
	\caption{Constraints on non-spinning primordial black hole dark matter with a monochromatic mass distribution over the entire viable mass range.  The various constraints are derived in this work (Super-K and INTEGRAL) and in Refs.\,\cite{Carr:2009jm, Boudaud:2018hqb, Niikura:2017zjd, Smyth:2019whb, Griest:2013esa, Allsman:2000kg, Tisserand:2006zx,Wyrzykowski:2011tr, Brandt:2016aco, Ali-Haimoud:2016mbv,Raidal:2017mfl,Authors:2019qbw,Zumalacarregui:2017qqd}.}
	\label{fig: Global plot}	
\end{figure*}

Fig.\,\ref{fig: Global plot} shows the constraints on non-spinning PBHs, with a monochromatic mass function, over the entire viable mass range.  We see that PBHs can form the entire DM density if it has a mass in the range of 2 $\times$ 10$^{17}$\,g -- 10$^{23}$\,g.  These constraints depend on the PBH mass distribution~\cite{Carr:2016drx,Carr:2017jsz,Green:2016xgy,Kuhnel:2017pwq,Bellomo:2017zsr}, however, it is generally true that there exist multiple mass distributions for which PBHs can make up the entire DM density.  Near future observations can completely probe this mass range\,\cite{Katz:2018zrn, Jung:2019fcs, Bai:2018bej,Munoz:2016tmg,Laha:2018zav}.

\end{document}